\documentclass{aa}  
\usepackage{graphicx}
\usepackage{txfonts}
\usepackage{xcolor}

\newcommand{\evh}[1]{{{#1}}}         
\newcommand{\bib}[1]{{{#1}}}         

\begin{document}

%
\newbox\grsign \setbox\grsign=\hbox{$>$} \newdimen\grdimen \grdimen=\ht\grsign
\newbox\simlessbox \newbox\simgreatbox
\setbox\simgreatbox=\hbox{\raise.5ex\hbox{$>$}\llap
     {\lower.5ex\hbox{$\sim$}}}\ht1=\grdimen\dp1=0pt
\setbox\simlessbox=\hbox{\raise.5ex\hbox{$<$}\llap
     {\lower.5ex\hbox{$\sim$}}}\ht2=\grdimen\dp2=0pt
\def\simgreat{\mathrel{\copy\simgreatbox}}
\def\simless{\mathrel{\copy\simlessbox}}
\newbox\simppropto
\setbox\simppropto=\hbox{\raise.5ex\hbox{$\sim$}\llap
     {\lower.5ex\hbox{$\propto$}}}\ht2=\grdimen\dp2=0pt
\def\simpropto{\mathrel{\copy\simppropto}}

\title{Another relic bulge globular cluster: ESO 456-SC38 (Djorgovski~2)\thanks{Observations obtained at the
Hubble Space Telescope, GO-14074 (PI: Cohen), 
and the European Southern Observatory, 
proposal 089.D-0493(A) (PI: Saviane).
} }

\author{
S. Ortolani\inst{1,2}
\and 
E. V. Held\inst{2}
\and
D. Nardiello\inst{1,2}
\and
S. O. Souza\inst{3}
\and
B. Barbuy\inst{3}
\and
A. P\'erez-Villegas\inst{3}
\and
S. Cassisi\inst{4,5}
\and
E. Bica\inst{6}
\and
Y. Momany\inst{2}
\and
I. Saviane\inst{7}
}
\offprints{B. Barbuy (b.barbuy@iag.usp.br).}

\institute{
Universit\`a di Padova, Dipartimento di Fisica e Astronomia Galileo Galilei, 
Vicolo dell'Osservatorio 2, I-35122 Padova, Italy
\and
INAF-Osservatorio Astronomico di Padova, Vicolo dell'Osservatorio 5,
I-35122 Padova, Italy
\and
Universidade de S\~ao Paulo, IAG, Rua do Mat\~ao 1226,
Cidade Universit\'aria, S\~ao Paulo 05508-900, Brazil
\and
INAF - Osservatorio Astronomico d’Abruzzo, via M. Maggini, sn. 64100, 
Teramo, Italy 
\and
INFN - Sezione di Pisa, Largo Pontecorvo 3, 56127 Pisa, Italy
\and
Universidade Federal do Rio Grande do Sul, Departamento de Astronomia,
CP 15051, Porto Alegre 91501-970, Brazil
\and
European Southern Observatory, Alonso de Cordova 3107, Santiago, Chile
}
 
\date{Received; accepted }
\abstract
{The object ESO456-SC38 (Djorgovski 2) is one of the globular clusters that is closest to 
the Galactic center. It is on the blue horizontal branch and has a moderate metallicity of [Fe/H]$\sim$-1.0. It is thus similar to the very old inner bulge
globular clusters NGC 6522, NGC 6558, and HP 1, and therefore appears to be part
 of  the primeval formation stages of    the Milky Way.}
{The aim of this work
 is to determine an accurate distance and metallicity for ESO456-SC38,
as well as orbital parameters, in order to
check similarities with other clusters in the inner bulge
 that have previously been well studied in terms of spectroscopy and photometry. This is
a considerably fainter cluster that is contaminated
by a rich stellar field; it is also quite absorbed by  the dusty foreground.}
{We analyzed ESO456-SC38 based on {{\it Hubble Space Telescope}} photometry,
with the filters F606W from ACS, F110W and F160W from WFC3, 
and photometry in V and I from FORS2 at the VLT. We combined this
with identified stars that are covered by Gaia Data Release 2.}
{ The isochrone fitting was carried out with the statistical Markov chain Monte Carlo method. 
We derive an accurate distance of d$_{\odot} = 8.75\pm 0.12$ kpc and a reddening of E(B-V)$=0.81^{+0.02}_{-0.02}$.
The best-fitting BaSTI isochrones correspond to an age of 
$12.70^{+0.72}_{-0.69}$ Gyr 
and a metallicity of [Fe/H]$=-1.11^{+0.03}_{-0.03}$.}
{ESO 456-SC38 adds to the list of moderately metal-poor globular clusters
located in the inner bulge. It is on the blue horizontal branch and is
very old.
The cluster is confined to the bulge and bar region, but it does not support
 the Galactic bar structure. The old stellar population represented by clusters like this
has to be taken into account in models of Galactic bulge formation. 
Studying them also
provides indications on the formation times of the globular clusters themselves. }
\titlerunning{Bulge globular cluster: ESO 456-SC38 (Djorgovski~2)}
\authorrunning{S. Ortolani et al.}
\maketitle
%

\section{Introduction}

 Colour-magnitude diagrams (CMD), abundances, and kinematics 
of globular clusters (GCs) have been  essential for understanding the formation  and  evolution of the Milky Way (MW). To mention a few, the studies of Searle \& Zinn (1978),  Renzini \& Fusi Pecci (1988),
Minniti (1995), Ortolani et al. (1995),
and Barbuy et al. (1998a), C\^ot\'e  (1999) at the end of the last century helped  shape the role of  GCs in our  
understanding of  the MW  as we see it today.
The multi-population content of many GCs is an additional 
clue to their very nature  (e.g., Piotto et al. 2015, and references therein).
 
 Globular clusters in the Galactic bulge consist of prominent  populous clusters
as well as 
some poorly populated  clusters. Minni 22  
is a recent example of the latter, with [Fe/H]~$ = -1.3$ 
(Minniti et al. 2018). On the other hand, the halo, in particular  
the outer halo, harbors  many faint and ultra-faint clusters (UFC)
that may  represent the low end of the luminosity function of GCs, 
at times at the verge of dissolution (see, e.g., Luque et al. 2016 for two
   faint halo clusters revealed 
in the Dark Energy Survey, DES). These halo UFCs are mostly old, with evidence that they have recently been 
accreted by the MW. Taking into account  faint bulge and halo stellar 
systems such as GC subclasses together with various GCs that are  extremely absorbed and/or 
contaminated by rich stellar fields,  Bica et al. (2019)
compiled 200 of them  in the MW  by including those studied by
Minniti et al. (2018) and others. This represents an increase 
of 37\% with respect to the compilation of Harris (1996), and is an increase of 27\% comparted to his 2010 version.  
Bica et al. also compiled 
94 GC candidates, most of  them toward the bulge. This suggests that additional 
GCs lie still farther  away in  the bulge region. 

The bulge  and bulge-halo (or inner halo) characterizations  can distinguish
 these
GC populations (Bica et al. 2016) as long as distances and other basic cluster 
parameters  are well constrained and   lead  to their orbits 
(e.g., P\'erez-Villegas et al. 2018, Ortolani et al. 2019).
The Galactic bulge contains a variety of clusters, from the
most metal-rich in the Galaxy to moderately metal-poor clusters
with a peak at [Fe/H]~$\sim -1.0$ (Bica et al. 2016).
Only one cluster is confirmed to be more metal poor than this,
which is Terzan 4 
with [Fe/H]~$\sim -1.6$ (Origlia \& Rich 2004).
 Bica et al. (2016) listed
the metallicities and abundances of bulge globular clusters obtained
with spectroscopy, that are available in the literature
(their Table 3). 
The compilation by Harris (1996, edition 2010) also includes estimates
based on photometry, which can be more uncertain.

There appears to be a concentration of GCs with a metallicity
of [Fe/H]$\sim -1.0$ together with a blue horizontal branch
(e.g., Bica et al. 2016, Cohen et al. 2018).
 Field halo stars with [Fe/H]$\sim -1.0$ are very rare because
this is the high end of the halo metallicity distribution function
(e.g., An et al. 2013);
 the distribution of globular cluster metallicities
from Harris (1996, edition 2010) shows a small peak at
[Fe/H]$\sim$-1.0, where essentially all
 correspond to clusters located in the bulge (Bica et al. 2016).

 These objects appear to be among the oldest  in  the central bulge and  the MW itself, such as NGC 6522, HP1, and NGC 6558.
These clusters were  analyzed spectroscopically for instance by 
Barbuy et al. (2014, 2016, 2018b).  NGC 6522 was found to be the
oldest in the sample by Meissner \& Weiss (2006), and also in
the estimation by Barbuy et al. (2009).
NGC 6522 and HP1 were shown to be very old, with about 13 Gyr, by 
Kerber et al. (2018, 2019). 
The old stellar populations seen in GCs  are thought to  
have a counterpart in the bulge field
(e.g., Renzini et al. 2018, and references therein). One particular
evidence for this are the RR Lyrae stars, which have a similar metallicity
peak at [Fe/H]$\sim$-1.0
as the moderately metal-poor bulge GCs (e.g., Pietrukowicz et al. 2012;
Saha et al. 2019). Further evidence for this comes from Schultheis et al. (2015), who have found a peak of bulge
field stars at [Fe/H]~$\sim-1.0$ that show an enhancement in alpha-elements,
and Schiavon et al. (2017), who identified a sample of nitrogen-rich
stars that also show metallicities of [Fe/H]~$\sim-1.0$.  

 Our group has been carrying out deep photometry and high-resolution 
spectroscopy, as well as proper-motion cleaned data, by employing
two epochs of bulge globular clusters
 (e.g., Rossi et al. 2015a, P\'erez-Villegas et al. 2018) to
derive their parameters and dynamical properties. The ultimate goal is
understanding the earliest formation processes of the central 
parts of the Galaxy (Barbuy et al. 2018a).

This study deeply explores an apparently  additional 
GC member of that group, ESO456-SC38 (Djorgovski 2). This object was already an entry in the Barbuy et al.  (1998) central bulge sample.  

The source ESO 456-SC38 was discovered by Holmberg et al. (1978)
 and published in  the sixth list of   
the ESO/Uppsala survey of the ESO (B) atlas of the southern sky. They classified
the object as an 
open cluster, while Djorgovski (1987) classified it as  a  GC, 
so that it  is often referred
to as Djorgovski 2  (or Djorg 2).

ESO456-SC38, or Djorgovski~2 (Djorgovski 1987),
 is located at $\alpha$ = 18$^{\rm h}$01$^{\rm m}$49.0$^{\rm s}$, 
$\delta$ = $-$27$^{\rm o}$$49'33\arcsec$, with Galactic coordinates
l = 2\fdg77, b = $-$2\fdg50.
ESO456-SC38 is projected  on a very contaminated central bulge stellar field. With CMDs in the 
$V$ and $I$ bands,
\bib{Ortolani et al. (1997)} derived  a reddening $E(B-V) = 0.85$ and a metallicity [Fe/H]~$ = -0.50$. A high metallicity, $[Fe/H]=-0.65$ dex, was also derived by Valenti et al (2010) from
  near-infrared photometry.
Harris (1996, version 2010) reported  $E(B-V) = 0.94$ and [Fe/H]~$ = -0.65$, and
also provided an integrated absolute magnitude of
$M_{V,t} = -0.70$. This means that this is an intermediate-luminosity GC. 

  V\'asquez et al. (2018)  derived a metallicity of  [Fe/H]~$\sim -1.1$ and a radial velocity of  v$_{\rm r} = -159.9 \pm 0.5$ km s$^{-1}$ from three stars
using spectroscopic observations of the \ion{Ca}{II} triplet (CaT).
Dias et al. (2016) obtained a heliocentric radial velocity of
  $-150 \pm 28$ km s$^{-1}$ using spectra in the visible region of four member
  stars. The derived metallicity is quite uncertain in the range
  [Fe/H] $=-0.5$ to $-1.19,$ depending on the adopted spectral library.

We here use accurate Gaia proper-motion selected CMDs for the bright
stars, combined with deep photometry from the Hubble Space Telescope (HST) and the Very Large Telescope (VLT),
to determine the metallicity and distance of this cluster. 
Orbital parameters are then derived in order to verify
its role as an additional  key GC for constraining  the bulge properties and its history.

In Section 2 we present  the observations and the data reductions. In Sect. 3
we derive the cluster parameters, based on HST and VLT CMDs
 with superimposed red giants that have been measured with Gaia. The orbit parameters are
 then derived in Sect. 4.
      Conclusions are drawn in   Sect. 5.
 

\section{Observations and data reduction}

\begin{figure}
\centering
\includegraphics[angle=0,width=1.0\columnwidth]{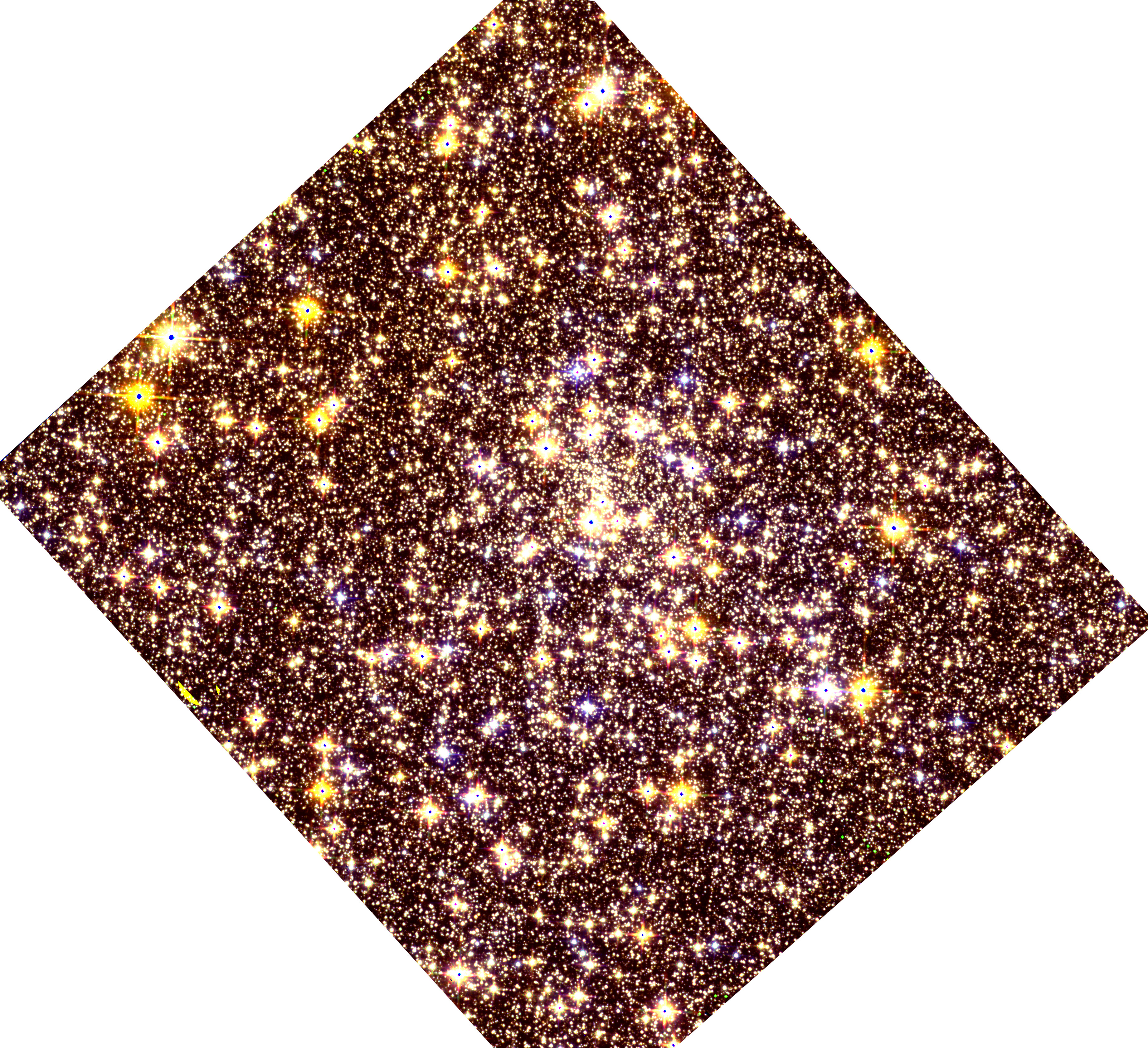}
\caption{Three-color image of the ESO456-SC38  field from HST data. We used the
  F606W, F110W, and F160W images for the blue, green, and red channel,
  respectively. North is at the top, and east to the left.
  The field size is $\sim 2.7$ arcmin left to right. }
  \label{dj2hst}
\end{figure}

We used available {HST} and {VLT} observations
  of the cluster.  We made use of {HST}
observations collected during the mission GO-14074 (PI:\,Cohen, mean
epoch 2016.4) in F606W with the Wide Field Channel (WFC) of the
Advanced Camera for Surveys (ACS), and in  F110W and F160W with the IR
channel of the Wide Field Camera 3 (WFC3).
For the data reduction we adopted the software \texttt{kitchen\_sync2},
developed by J.~Anderson (Anderson et al. 2008) that was also used by 
Nardiello et al. (2018) to extract an atlas of 56 GCs. 
Briefly, using
\texttt{flc} images 
(i.e., corrected for charge transfer inefficency, Anderson \& Bedin 2010)
and perturbed empirical arrays of point spread functions (PSF),
the software analyzes all the exposures simultaneously to find and measure the
sources. We refer to
Nardiello et al. (2018) and Bellini et al. (2017)
 for a detailed description of this approach.
The magnitudes were calibrated into the Vega-mag system by comparing
aperture photometry on \texttt{drc} images against our PSF-fitting
photometry and adopting the photometric zero-points given by the ACS
zero-points calculator\footnote{https://acszeropoints.stsci.edu/} in
the case of ACS/WFC observations, and by \bib{Kalirai et al. (2009)}
in the case of WFC3/IR data.
A three-color image of the field observed with {HST} is shown in Fig.~\ref{dj2hst}.

\begin{figure}
\centering
  \includegraphics[angle=0,width=0.7\columnwidth]{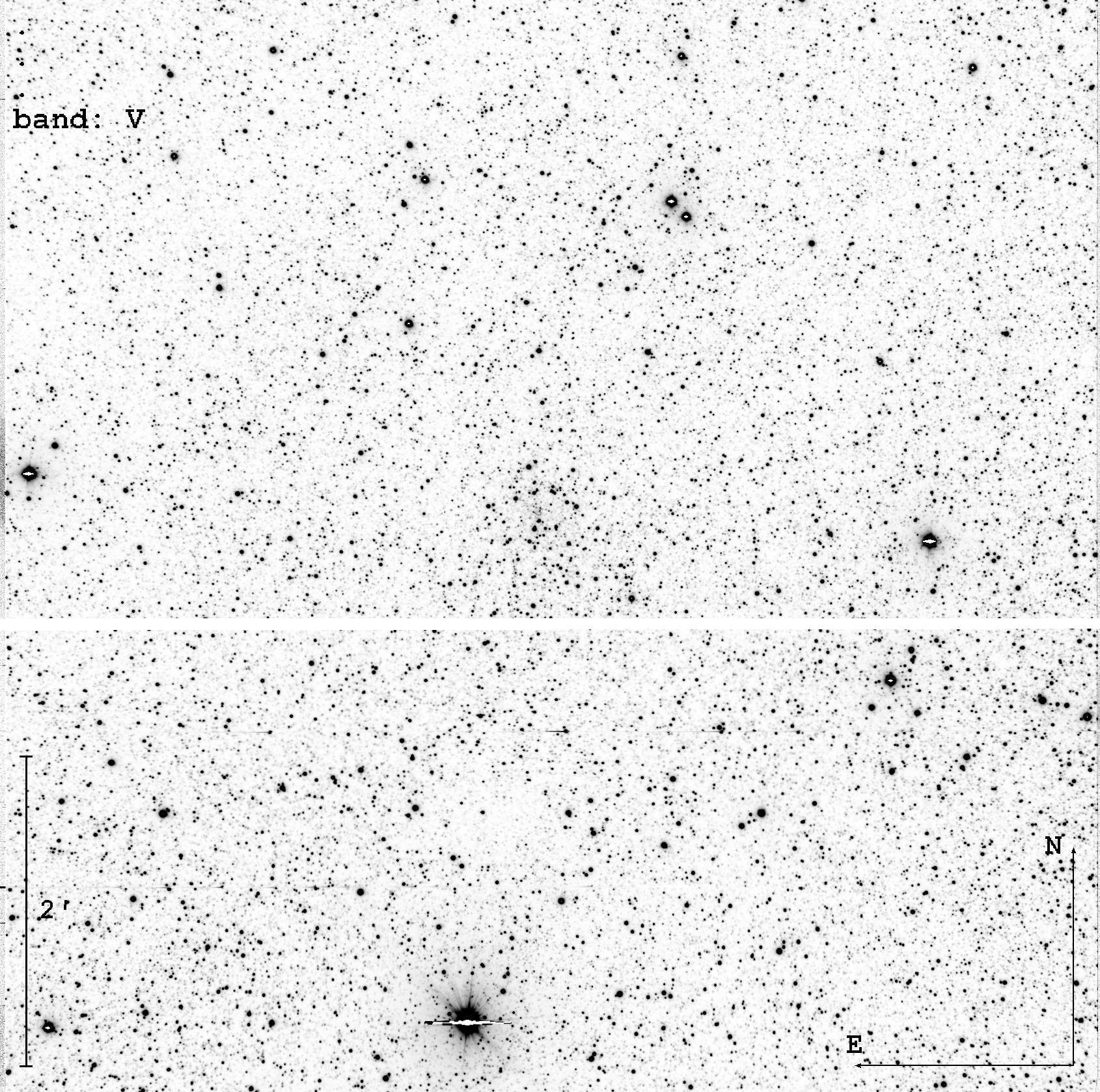}
 \caption {FORS2 $V$ image of ESO456-SC38. North is up, and east to the left.
 The field size is 7$\times$7 arcmin$^{2}$. }
 \label{dj2fors}
\end{figure}

The VLT imaging observations were obtained with FORS2 during
  a preimaging run for our spectroscopic survey of MW globular clusters
(Saviane et al. 2012; Dias et al. 2016; V\'asquez et al. 2018).  The
  cluster was observed under excellent seeing conditions ($\sim0.5$
  arcsec FWHM), using 30s exposures in $V$ and $I$ complemented by
  very short exposures (0.25s) to allow photometry of the otherwise
  saturated brightest red giant stars.  Photometry was obtained with
  \texttt{daophot} and \texttt{allstar} (Stetson 1987).  Because sky
  conditions were not always photometric during the preimaging
  observations, the zero-point of our photometry was tied to the
  well-calibrated observations obtained at the Danish telescope by
  Ortolani et al. (1997), properly taking into account the color
  transformations.  The original zero-point error was estimated
to be of 0.03 mag.  Because the transfer error is
0.02 mag, we estimate a total budget error of 0.04 mag for both colors.
 In Fig.~\ref{dj2fors}  we show a $V$ image of
  ESO456-SC38 from FORS2.

\section{Color-magnitude diagrams and cluster parameters}

\begin{table}
\centering
\caption[1]{\label{literature}Literature values of ESO~456-SC38. References: 1. Ortolani et al. (1997); 2 Harris (1996, updated in 2010);
 3 Valenti et al. (2010); 4 V\'asquez et al. (2018).}
\begin{tabular}{@{}cccccc@{}}
\noalign{\smallskip}
\hline
\noalign{\smallskip}
  {\rm \phantom{-}\phantom{-}E(B-V)} & d$_{\odot}$
& {\rm R$_{\rm GC}$} & {\rm [Fe/H]} 
& v$_{\rm r}$ & ref.  \\
& (kpc)& (kpc)& (dex) & (km s$^{-1}$)& \\
\noalign{\smallskip}
\hline
\noalign{\smallskip}
$0.89\pm0.08$ & $5.5\pm0.8$    & $2.5$         & $-0.5$     & ---     &  1 \\
$0.94\pm0.15$       & $6.3\pm1.0$    & $1.8$         & $-0.65$     &  ---    & 2 \\
 $0.94\pm0.15$ & $7.0$         & $1.1$         & $-0.65$     & ---    & 3 \\
 ---           & ---          & ---      & $-1.07\pm0.21$ & $-159.9$ & 4 \\
\noalign{\smallskip}
\hline
\end{tabular}
\end{table}

\subsection{Gaia DR2 cross-identifications}

The merged catalogs of ACS+WFC3 and FORS2 at the VLT were matched to the Second
 Data Release of Gaia (Gaia collaboration 2018a,b)
 using \texttt{topcat} (Taylor 2005). 
 There are only a few stars in common, therefore no cuts due to Gaia DR2
flags were applied.
For the match we adopted a maximum distance  of 0.3 arcsec.  A plot of the cross-identified stars in the proper motion space 
shows a clearly detectable excess around $\mu_\alpha^* = 0.52$ mas/yr, $\mu_\delta = -3.15$ mas/yr, which we identified as the mean proper motion of ESO456-SC38,  in agreement with the value derived by 
Vasiliev (2019) and Baumgardt et al. (2019). We selected a subsample of stars within a radius of 1.7 mas around the mean proper motion
\evh{to increase the cluster membership probability and reduce the  contamination} by the Galactic disk and bulge. The mean parallax of this subsample is 0.075 mas ($\sigma_{\rm plx}= 1.02$). In view of this large error, we refrain from deriving a geometric distance from this parallax. However, we note that adding the mean systematic error that is known to affect the parallax in Gaia DR2
(Gaia Collaboration et al. 2018b), this value suggests  a long distance of $\sim 9.6$ kpc.

\begin{figure*}
\centering
\includegraphics[width=2\columnwidth]{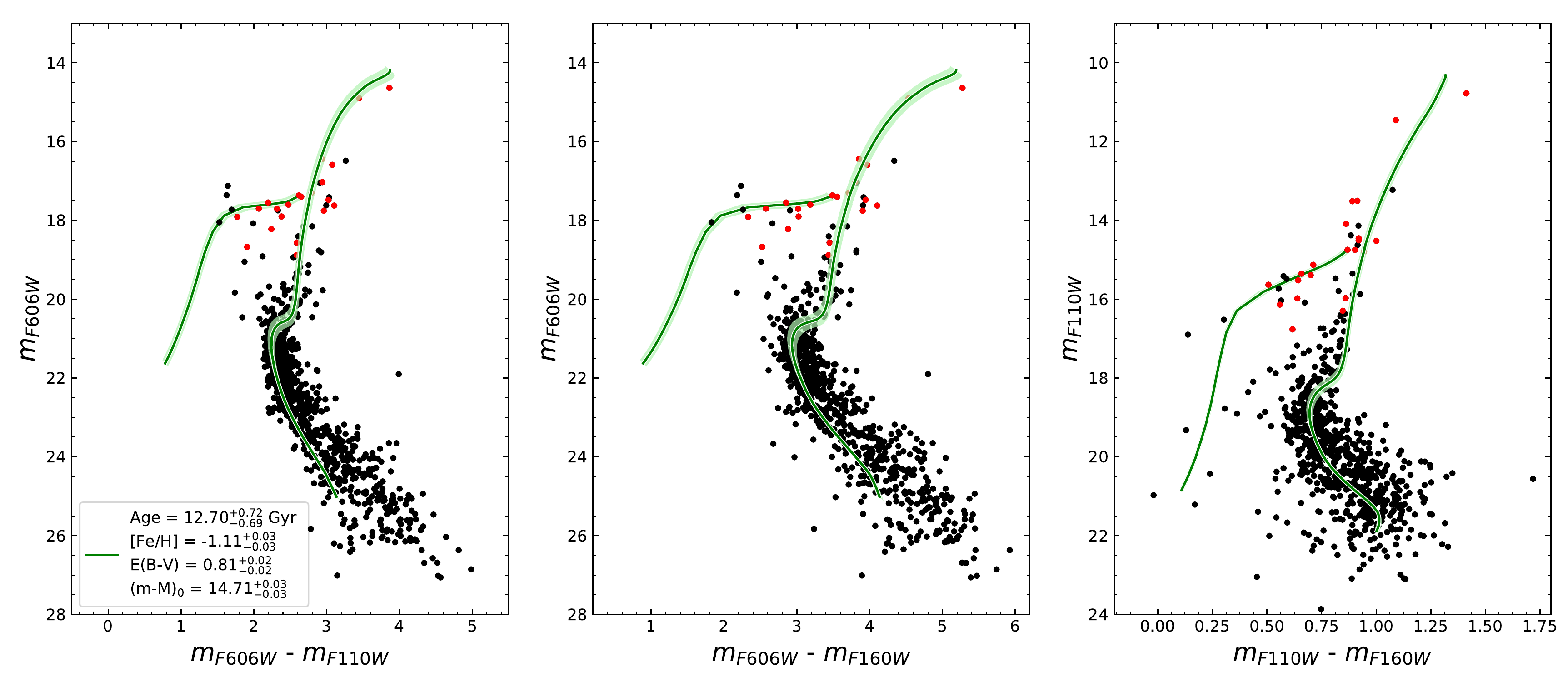}
\caption{ Color-magnitude diagram  m$_{F606W}$ vs. m$_{F606W}$-m$_{F110W}$ (left panel) of ESO456-SC38 from HST/ACS and
 WFC3 observations fit with the statistical Bayesian MCMC method (SIRIUS code), based on a grid of BaSTI isochrones.
 The green solid lines are the most probable solution, and the green region represents the solutions inside 1$\sigma$ from
the posterior distributions. Large red dots are stars identified in Gaia within 25 arcsec of the cluster center. The middle
and right panels show the solution computed in the left panel projected over the m$_{F606W}$ vs. m$_{F606W}$-m$_{F160W}$ (middle)
and m$_{F110W}$ vs. m$_{F110W}$-m$_{F160W}$ (right) CMDs.
 Stars within 0.3 arcmin of the cluster center are selected.}
\label{stefano}
\end{figure*}

\begin{figure}
\centering
\includegraphics[viewport= 0 0 490 520, clip=true, width=0.9\columnwidth]{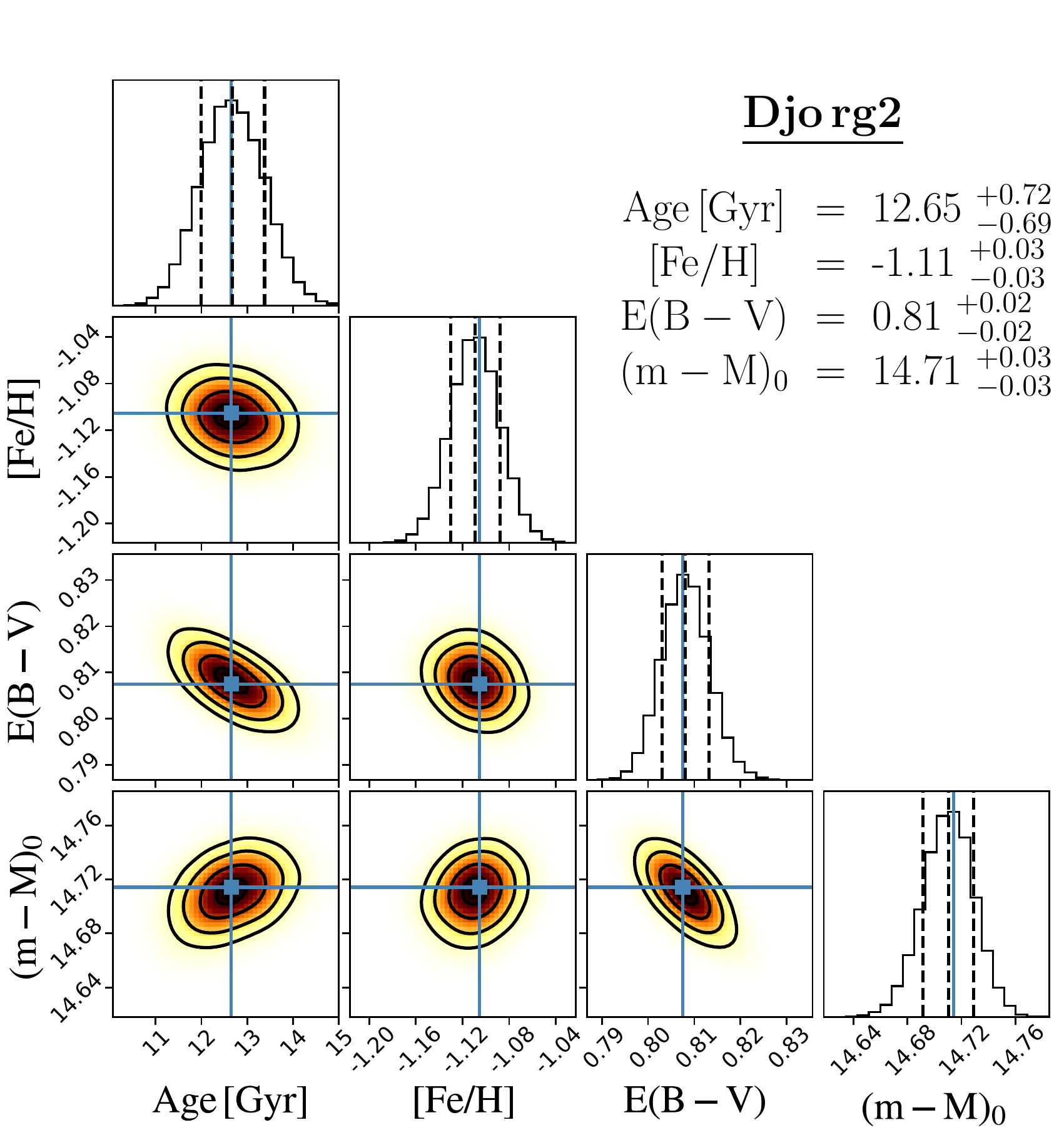}
\caption{Corner plot relative to the  m$_{F606W}$ vs. m$_{F606W}$-m$_{F110W}$
CMD in Fig. \ref{stefano} (left panel). }
\label{cornerplot}
\end{figure}

\subsection{Color-magnitude diagrams and isochrone fitting}

\evh{In Figs.~\ref{stefano} and~\ref{forscmd+rrl} we plot the 
inner region of the CMD
(within 0.3 arcmin and 0.9 arcmin of the cluster center, respectively) 
from the HST and VLT data.} 
The subsample of probable cluster members from the Gaia DR2 match is plotted as red dots. 

The BaSTI (Pietrinferni et al. 2004, 2006) alpha-enhanced stellar
evolution models\footnote{http://basti-ao-abruzzo.inaf.it/index.html} 
 were adopted for the isochrone fitting.
The isochrones were corrected for reddening dependent on effective temperatures, as discussed in
Ortolani et al. (2017). We used for that purpose the star-by-star reddening provided for the Padova isochrone model  set\footnote{http://stev.oapd.inaf.it/cgi-bin/cmd\_2.3}
\bib{(Bressan et al. 2012)}.
This correction mainly shrinks the CMDs in color,  and the fit quality greatly improves for very reddened clusters, as shown for 
Djorg~1 (Ortolani et al. 2019).


 In Fig. \ref{stefano} the HST m$_{F606W}$ vs.
  m$_{F606W}$ - m$_{F110W}$ CMD
of ESO46-SC38 is fit with a Bayesian approach. 
The code Statistical Inference of physical paRameters of
sIngle and mUltiple populations in Stellar clusters (SIRIUS) by
Souza et al. (in preparation) applies
 statistical isochrone fitting following a Bayesian approach
with the Markov chain Monte Carlo (MCMC) sampling method, in order 
to derive the age, metallicity, reddening, and distance.
 The method is described in Kerber et al. (2019).

In the middle and right panels of Fig. \ref{stefano} we superimpose 
the same parameters as are adopted in the left panel of Fig. \ref{stefano} for the CMDs 
 m$_{F110W}$ vs.
  m$_{F110W}$ - m$_{F160W}$ (middle) and  m$_{F606W}$ vs.
  m$_{F606W}$ - m$_{F160W}$ (right). This shows a very good compatibility.
A corner plot relative  to the  m$_{F606W}$ vs.
  m$_{F606W}$ - m$_{F110W}$ CMD (left panel in Fig. \ref{stefano}) is presented
in Fig. \ref{cornerplot}.

In these figures, the isochrone fitting the deep HST photometry 
is located near the blue edge of the main sequence (MS). This is due to a combined
effect of field contamination (prevailing on the red side), binarity,
and loss of completeness in the cluster sequence at the faint end. 

We also point out that the presence of a 
blue horizontal branch (BHB) is confirmed from counting
the relative number of stars in the horizontal branch (HB) and
red giant branch (RGB). The number of HB
stars is about 20-30\% higher
than the number of RGB stars, in agreement with other globular clusters and the
typical evolution timescale ratio (see, e.g., Iben et al. 1969).
In addition, the BHB is well fit  in Fig. \ref{stefano}.
The foreground and  background contamination from the MW  is small 
in the region of the HB.

 By adopting this statistical method for the fitting procedure,
 we derive an age of $12.70^{+0.72}_{-0.69}$ Gyr, a metallicity of
[Fe/H]$=-1.11^{+0.03}_{-0.03}$, a reddening of E(B-V)$=0.81^{+0.02}_{-0.02}$ and
a distance modulus of (m-M)$_{0} = 14.71^{+0.03}_{-0.03}$ or a distance of $8.75\pm0.12$ kpc.
The metallicity obtained from the MCMC fitting of 
[Fe/H]=-1.11$\pm$0.03 and [$\alpha$/Fe]=+0.4 (Z$\approx$0.004)
is in excellent agreement
with the spectroscopic value of [Fe/H]~$=-1.09$ by V\'asquez et al. (2018).

\subsection{Distance to ESO456-SC38}

\begin{figure}[]
\centering
\includegraphics[viewport= 0 130 490 720, clip=true, width=1.0\columnwidth]{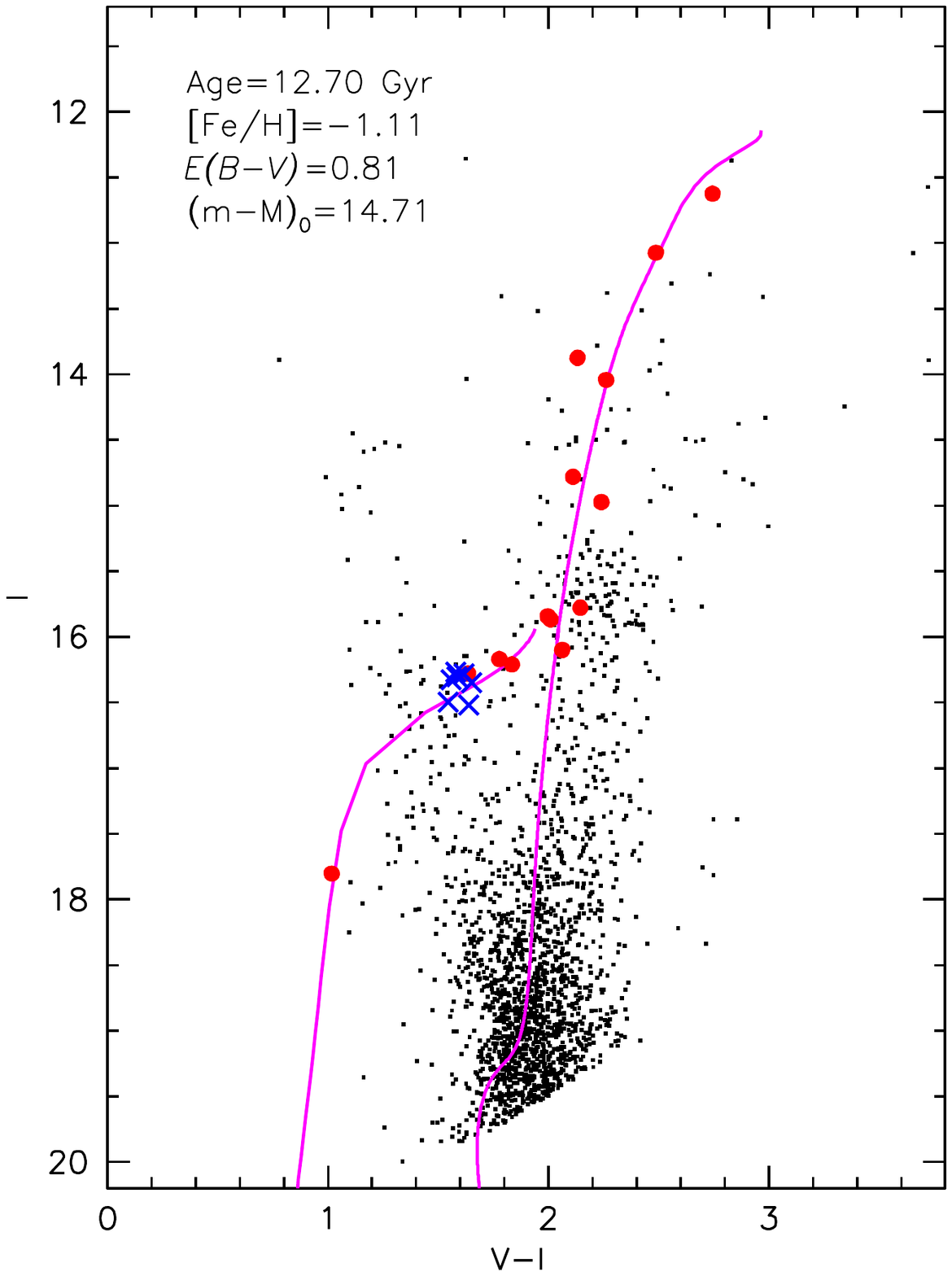}
\caption{$I$ vs. $V-I$ CMD from FORS2, overimposed with a BaSTi isochrone
of 12.7 Gyr and [Fe/H]=$-$1.1 (Z=0.004) (magenta line). 
 Stars within 0.9 arcmin of cluster center are selected.
The  mean magnitudes and colors of  7 RR Lyrae are shown with blue crosses.
Black dots are stars within 18 arcsec from the cluster center. 
Large red dots are stars
identified in Gaia within 25 arcsec of cluster center. }
 \label{forscmd+rrl}
  \end{figure}

 In Fig. \ref{forscmd+rrl} we compare the FORS2 data with a BaSTI isochrone using the
parameters derived from the MCMC isochrone fit to HST data. This
comparison provides support and confirmation of the HST fit.

We inspected the 
OGLE III catalog of RR Lyrae stars (Soszy{\'n}ski  et al. 2014) 
and considered seven of them that are also reported in Clement et al. (2001).
These seven RR Lyrae stars  (plotted as crosses in Fig.~\ref{forscmd+rrl}) 
have mean  magnitudes  $\langle I \rangle =16.35$, $\langle V \rangle = 17.954$,   and $V-I = 17.954 - 16.35 = 1.60$. 
Assuming $M_{V} = 0.214 \times {\rm [Fe/H]}+0.88^{+0.04}_{-0.06}$ (Gaia Collaboration et al. 2017)
and a metallicity [Fe/H]~$= -1.11$, we obtain  $M_{V}  = 0.64$. 
 The error in $M_{V}$ comes
from the error in metallicity, which in our case is $\pm$0.03.
Assuming total-to-selective absorption $R_{V} = 3.1$, and $E(B-V) = 0.81$,
 the distance modulus from RR Lyrae results in
(m-M)$_{0}$ = 14.8$^{+0.07}_{-0.09}$ and a distance of  9.12$^{+0.30}_{-0.37}$ kpc.
This distance is  in good agreement with the
distance values from the CMD fits, 8.75$\pm$0.12.
This distance is significantly larger than the 6.3 kpc value given 
in Harris (1996)
that was adopted by Baumgardt et al. (2019) for their dynamical calculations.

\begin{figure}[]
\centering
\includegraphics[viewport= 0 130 690 720, clip=true, width=1.0\columnwidth]{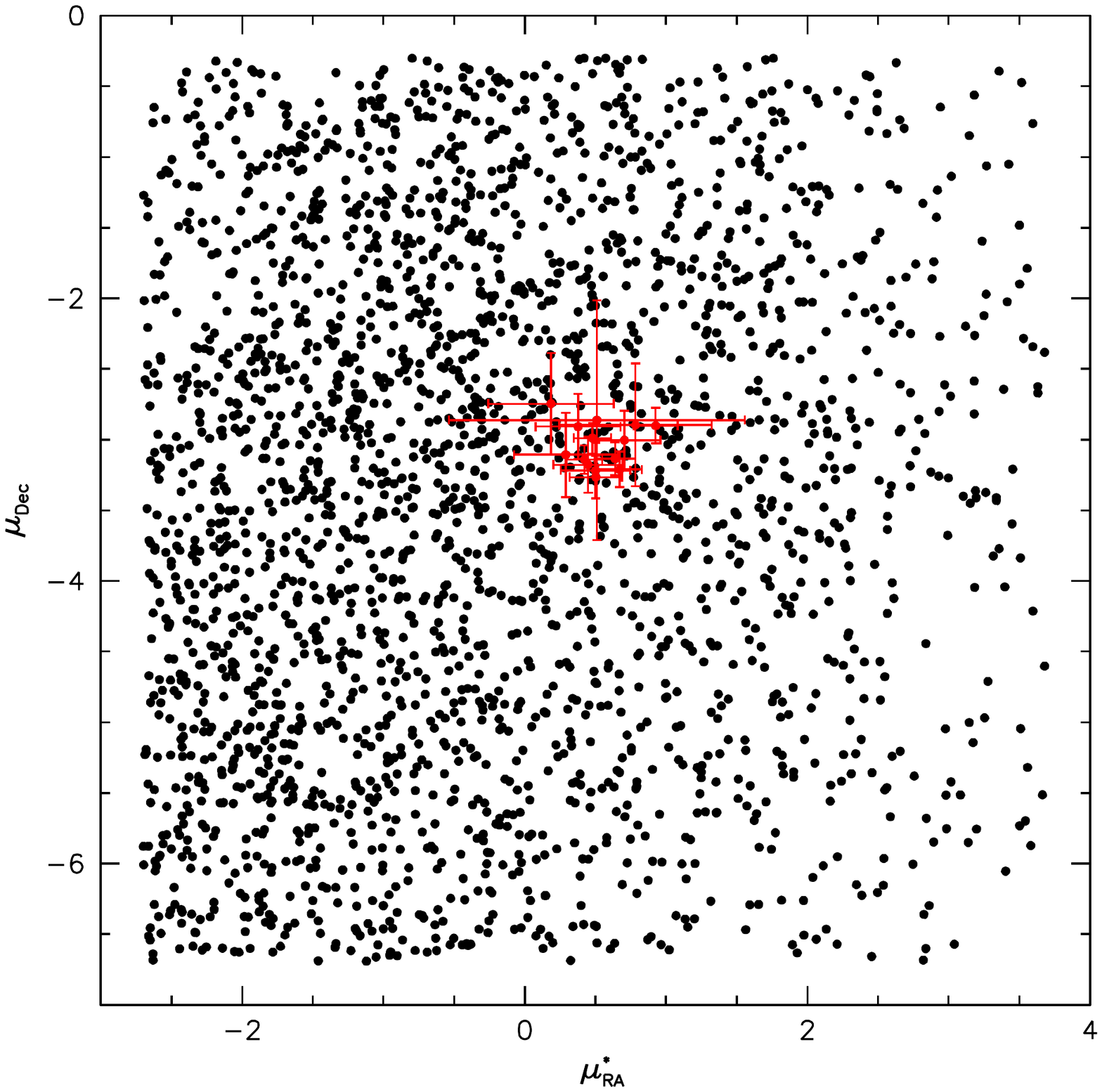}
\caption{Distribution of  Gaia-identified stars in proper
motion space, with errors given in  Gaia DR2.}
 \label{pmplot}
  \end{figure}

 For the data in Fig. \ref{forscmd+rrl}, we adopted a radial selection within 0.9
arcmin, which is the scale radius of a Plummer profile from Vasiliev
(2019).  For our stars in Gaia DR2, the sample is defined by a more
restrictive radial selection of half of the scale radius plus a
selection in proper motion space (vector diagram) equal to a radius
0.5 mas/yr around the mean proper motion of the cluster as determined
by Vasiliev (2019) and Baumgardt et al. (2019) using a rigorous
statistical approach.  The radius was selected by visual inspection of
the diagram to restrictively encompass the cluster overdensity in the
proper motion space. This region is comparable to the region of the cluster
member candidates described by Vasiliev (2019) in his Figure 1.
The distribution of our Gaia-identified star sample in proper
motion space is shown in Fig. \ref{pmplot} with errors given in Gaia DR2.

\section{Orbital properties}

For the first time, we have the entire information on ESO456-SC38 that is needed to estimate its probable Galactic orbit. We used a combination of the proper motions from Gaia DR2 (from Vasiliev 2019, confirmed by our analysis),
the recent radial velocity determinations using CaT spectra  (V\'asquez et al. 2018), and the accurate distance calculated in this work.  

In our analysis, we employed a nonaxisymmetric model for the Galactic gravitational potential. The model has an axisymmetric background made by an exponential disk built from the superposition of three Miyamoto-Nagai potentials (Miyamoto \& Nagai 1975) following the recipe by
Smith et al. (2015), and a Navarro-Frenk-White (NFW) density profile
(Navarro et al. 1997) to model the dark matter halo, which has a circular velocity $V_0=241$ km s$^{-1}$ at $R_0=8.2$ kpc \bib{(Bland-Hawthorn \& Gerhard 2016)}. On the axisymmetric background, a triaxial Ferrer's bar potential is superimposed. The total bar mass is $1.2 \times 10^{10}$ M$_{\odot}$, an angle of $25^{\circ}$ with the Sun-major axis of the bar, a gradient of the pattern speed of the bar of $\Omega_b= 40$, 45, and 50 km s$^{-1}$ kpc$^{-1}$, and a major axis extension of 3.5 kpc. We kept the same bar extension, but changed the bar pattern speed. 

The orbits were integrated with the \texttt{NIGO} tool 
(Rossi 2015b), which includes the potentials mentioned above. The solution of the equations of motion was found numerically using the Shampine--Gordon algorithm (for details, see Rossi 2015c). The initial conditions of ESO456-SC38 were obtained from the observational data, coordinates, heliocentric distance, radial velocity, and absolute proper motions given in Table~\ref{tab:paraGC}. The velocity components of the Sun with respect to the local standard of rest are $(U,V,W)_{\odot}= (11.1, 12.24, 7.25)$ km s$^{-1}$ \bib{(Sch{\"o}nrich et al. 2010),  where U, V, W, are positive in the direction of the Galactic center, Galactic rotation, and North Galactic Pole, respectively}. We generated a set of 1000 initial conditions using a Monte Carlo method, taking into account the errors in distance, heliocentric radial velocity, and absolute proper motion in both components to estimate the effect of the uncertainties associated with the cluster parameters. We integrated these initial conditions forward for 10 Gyr. 

For each orbit, we calculated the perigalactic distance $r_{\rm min}$, apogalactic distance $r_{\rm max}$, the maximum vertical excursion from the Galactic plane $|z|_{\rm max}$, and the eccentricity defined by $e=(r_{\rm max}-r_{\rm min})/(r_{\rm max}+r_{\rm min})$. In order to determine whether the orbital motion of the cluster has a prograde or a retrograde sense with respect to the rotation of the bar, we calculated the $z-$component of the angular momentum in the inertial frame, $Lz$. Because this quantity is no longer conserved in a model with nonaxisymmetric structures, we are interested only in the sign of the maximum and minimum $Lz$.

\begin{table}
\caption{Parameters for the orbit integration.}
\begin{tabular}{@{}lcc@{}} 
\hline
Parameter & Value & Ref. \\
\hline
$\alpha,\delta\ {(\rm J2000)}$ & 18$^{\rm h}$01$^{\rm m}$49.0$^{\rm s}$, $-$27$^{\rm o}$$49'33\arcsec$ & This work  \\
$V_r$ (km s$^{-1}$)& $159.9\pm0.5$ & 1  \\
$d_\odot$ (kpc) &$8.75 \pm 0.12$ & This work  \\
$^{\dag}$$\mu_{\alpha} \cos \delta$ (mas yr$^{-1}$) &$0.576\pm 0.060$ & 2 \\
$^{\dag}$$\mu_{\delta}$ (mas yr$^{-1})$& $-3.048\pm 0.055$ & 2 \\
\hline
\end{tabular}
\label{tab:paraGC}
\\References --- (1) V\'asquez et al. (2018); (2) Vasiliev (2019). 
\\$^{\dag}$Uncertainty includes a systematic error of 0.035 mas yr$^{-1}$
(Gaia Collaboration et al. 2018b). 
\end{table}

\begin{figure*}
\centering
\includegraphics[angle=0,width=1.0\columnwidth]{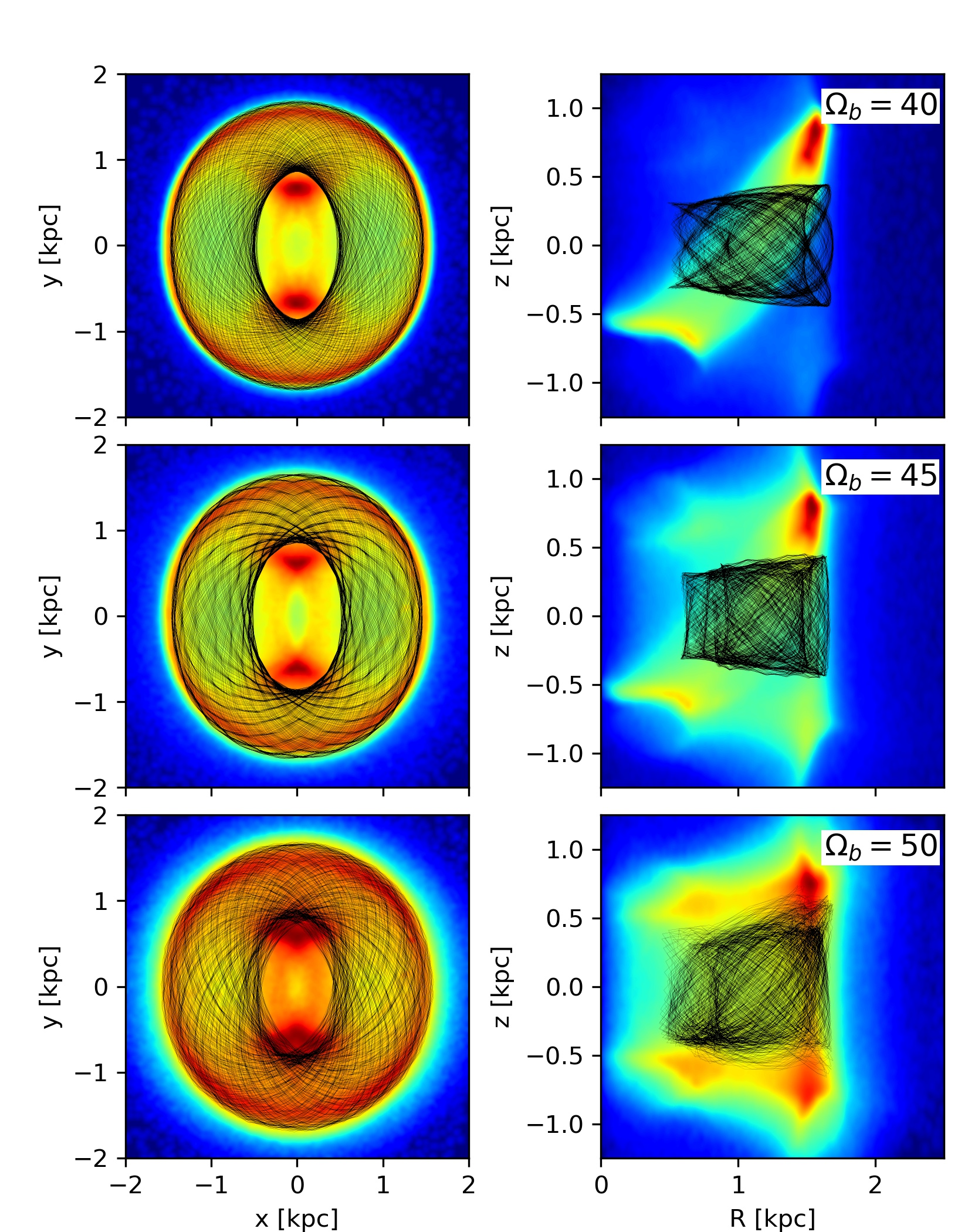}
\caption{Probability density map for the $x-y$ and $R-z$ projections of the 1000 orbits for ESO456-SC38. The orbits corotate with the bar frame. The red and yellow colors correspond to the higher probabilities. The black lines show the orbits using the central values presented in Table~\ref{tab:paraGC}.}
  \label{orbits}
  \end{figure*}

Figure \ref{orbits} shows the probability densities of the orbits in the $x-y$ and $R-z$ projection co-rotating with the bar. The red and yellow colors exhibit the region of space that the set of orbits for ESO456-SC38 crosses most frequently. The black curves are the corresponding orbits using the central values of the cluster observational parameters given in Table \ref{tab:paraGC}.

\begin{figure*}
\centering
\includegraphics[angle=0,width=2.0\columnwidth]{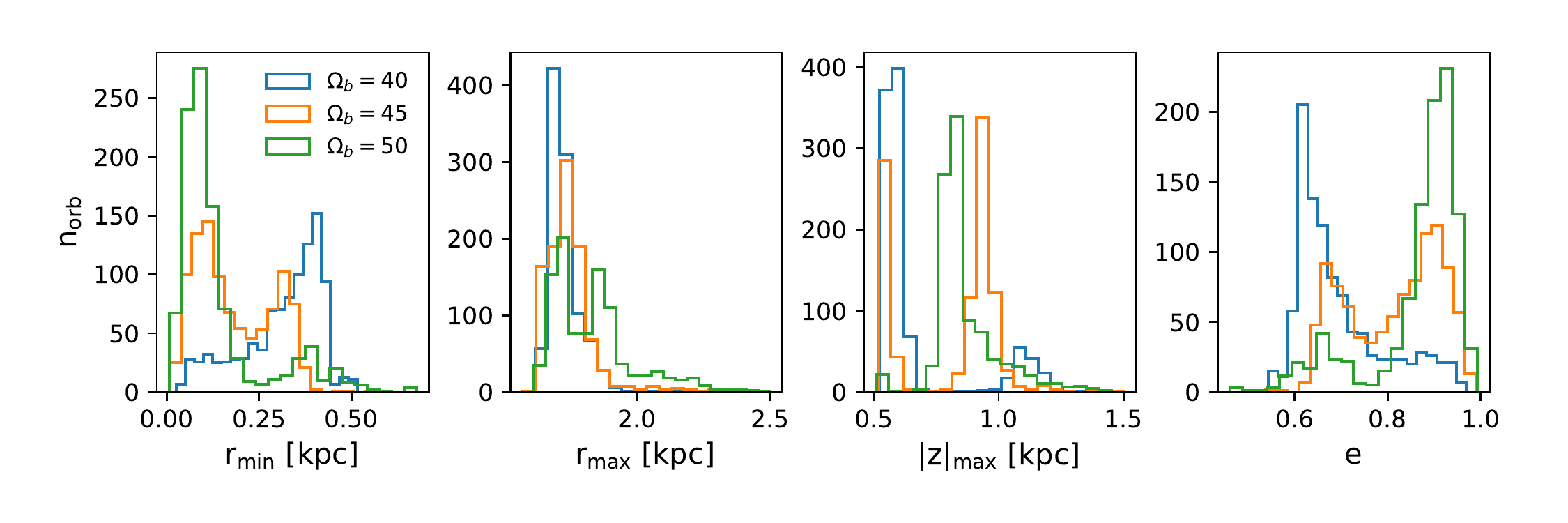}
\caption{Distribution of orbital parameters for ESO456-SC38, perigalactic distance $r_{\rm min}$, apogalactic distance $r_{\rm max}$, maximum vertical excursion from the Galactic plane $|z|_{\rm max}$, and eccentricity. The colors show the different angular speed of the bar, $\Omega_b$ = 40  (blue), 45 (orange), and 50 (green) km s$^{-1}$ kpc$^{-1}$.}
  \label{histogram}
  \end{figure*}

\begin{figure*}
\centering
\includegraphics[angle=0,width=2.0\columnwidth]{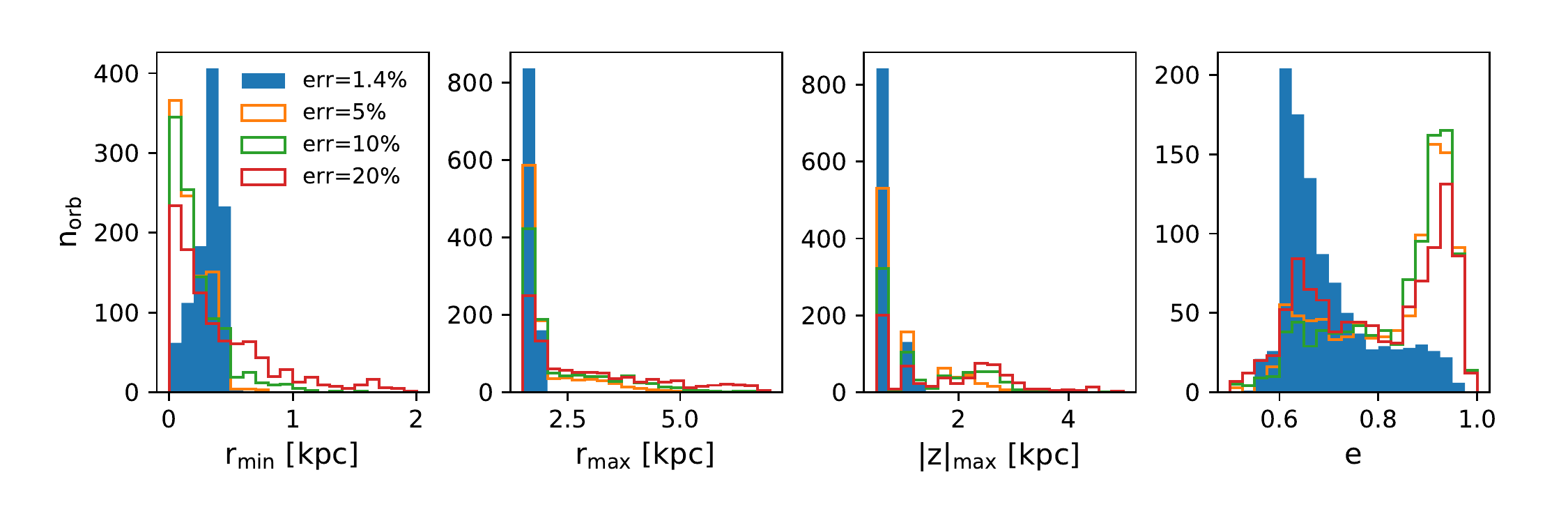}
\caption{Same as Fig. \ref{histogram}, but comparing the distribution of orbits as assumed in this paper,
and larger errors in distance (see text), computed with a pattern speed of 40 kms$^{-1}$ kpc$^{-1}$ .}
  \label{histogramerror}
  \end{figure*}

Distributions for the perigalactic distance, apogalactic distance, maximum vertical height, and eccentricity for ESO456-SC38 are presented in Fig.~\ref{histogram}, where the different colors represent the angular velocities investigated here.   The cluster has radial excursions between $ \sim 0.1$ and $ \sim 2.5$ kpc with maximum vertical excursions from the Galactic plane between $\sim 0.5$ and $\sim1.5$ kpc, and the eccentricity is between $ \sim 0.6$ and $\sim 0.9$. Additionally, the distribution of the perigalactic distance and eccentricity are sensitive to the pattern speed, that is, with a faster angular velocity of the bar, the perigalactic distance is closer to the Galactic center and the eccentricity increased. In the maximum vertical excursion, the orbits are more confined to the plane with a lower pattern speed. The apogalactic distance behaves similarly as the pattern speeds. Even though the sample of orbits for ESO456-SC38 is confined inside 2.5 kpc from the Galactic center, the orbits do not support the bar structure. Most of the orbits are prograde and retrograde at the same time when the pattern speed is slower, while most of the orbits are retrograde when the pattern speed is faster, but there are no prograde orbits.
 Prograde-retrograde orbits have been observed in other globular clusters
\bib{(Pichardo et al. 2004; Allen et al. 2006)}. This effect is produced by the bar, and it could be related to chaotic behavior, but this is not yet well understood.
In particular, the orbit calculated with the central values of the observational parameters is retrograde for the three different pattern speeds. Table \ref{tab:orbital} gives the average of the orbital parameters for ESO456-SC38. The errors provided in each column are the standard deviation of the distribution.

\begin{table*}
\caption[1]{Orbital parameters for ESO456-SC38 for three pattern speeds of the bar.}
\label{tab:orbital}
\begin{flushleft}
\begin{tabular}{@{}ccccc@{}}
\noalign{\smallskip}
\hline
\noalign{\smallskip}
$\Omega_b$ & $\langle r_{\rm min}\rangle$ & $\langle r_{\rm max} \rangle$ & $\langle |z|_{\rm max}\rangle $ &$\langle e\rangle$ \\
(km s$^{-1}$ kpc$^{-1}$) & (kpc) & (kpc) & (kpc) & \\
\noalign{\smallskip}
\hline
\noalign{\smallskip}
40 &    0.32  $\pm$ 0.11 & 1.73 $\pm$0.07 & 0.67 $\pm$ 0.20 & 0.70 $\pm$ 0.10 \\
45 & 0.18 $\pm$0.10 & 1.75 $\pm$ 0.12 &  0.82 $\pm$ 0.21 & 0.81 $\pm$ 0.10 \\ 
50 & 0.14 $\pm$ 0.12 & 1.84 $\pm$ 0.16 & 0.87 $\pm$ 0.14 & 0.87 $\pm$0.10 
  \\
\noalign{\smallskip} \hline \end{tabular}
\end{flushleft} 
\end{table*}

\emph{Uncertainties in orbits:}
 The range of orbital parameters can be affected by the distance uncertainty.
  Because we performed a statistical isochrone 
fitting where we took the photometric uncertainties given by the HST into account using information of the horizontal branch and metallicity, we trust our distance determination with an uncertainty
 of 0.12 kpc, which is equivalent to 1.4\%.  
To show how much the range of the orbital parameters would change, we calculated three sets of orbits, assuming 5, 10, and 20\% 
in the distance error. As expected, the parameter range increases with the error.
As shown in Fig. \ref{histogramerror}, with an error larger than 5\%, 
the cluster could reach distances out of the bulge and bar region and vertical excursions from the plane higher than 2 kpc,
 which would cause more eccentric orbits. Additionally, with larger uncertainties, the cluster is freer to reach regions of the Galaxy
 that would not be accessible with a smaller uncertainty. In other words, it could be closer or farther to some resonance that 
could affect its orbit. With an uncertainty of 0.12 kpc, most of the orbits are retrograde (for the slower pattern speed of
 the bar) with no prograde orbits, but when the error increases, the cluster would present some prograde orbits, and  
their number increases with distance error. 
For this reason, it is crucial to have the best possible accurate distance determination for the clusters in the bulge,
in order to be able to construct precise orbits around the Galaxy.

\section{Summary and conclusions}

We have gathered photometric data from the Hubble Space Telescope
in F606W with the ACS camera, and in F110W with the WFC3 camera,
together with images in V and I obtained with FORS2 at the Very
Large Telescope. These data were further cross-identified with Gaia
measurements of proper motions. This combination of data allowed
us to derive an accurate distance of d$_{\odot}$ = $8.75\pm0.12$.
The best-fitting BaSTi isochrones correspond to an age of $12.70^{+0.72}_{-0.69}$ Gyr,
a reddening of E(B-V)$=0.81\pm0.02$, and a metallicity
of [Fe/H]$=-1.11\pm0.03$.

 The source ESO456-SC38 is an addition to the list of moderately metal-poor
 ([Fe/H]$\sim- 1.0$) clusters on the blue horizontal branch, and
is projected very close to the Galactic center.
It appears to be similar to
NGC~6522 (Barbuy et al. 2014, Kerber et al. 2018),
 HP~1 (Barbuy et al. 2016, Kerber et al. 2019),
and NGC~6558 (Barbuy et al. 2018b).
The Galactic bulge currently appears to be dominated by a bar, but
in its early beginnings, a classical bulge was probably present, as
proven by the concentration of these very old GCs. This is further
shown by the presence of RR Lyrae stars that also have a peak in metallicity at [Fe/H]$\sim$$-$1.0 (Pietrukowicz et al. 2012).
The scenario could be an early small classical bulge,
or else a very early thick disk, with
a fast chemical enrichment reaching [Fe/H]$\sim -1.3$ to $-$1.0, 
when most metal-poor stars and GCs formed. 
The subsequent evolution of the bulge
is dominated by the effects of the bar. For example,
Zoccali et al. (2018) measured
 red clump stars and suggested that the bulge
currently contains two main components, a metal-poor component with a peak at
[Fe/H]$\sim$-0.4, and a metal-rich one with a peak at [Fe/H]$\sim$+0.3.
In addition to the difference in metallicity, these components also show different spatial distribution, kinematics,
and abundance ratios.  The metal-rich
end of the metallicity distribution function in the bulge varies from one study to another
(e.g., Barbuy et al. 2018a). We note for metal-poor stars
  that the surveys that are based on red clump stars are biased against
metal-poor stars. For the detection of the old stars, surveys of
RR Lyrae stars are suitable (Saha et al. 2019; Pietrukowicz et al. 2012),
and spectroscopic surveys such as
Extremely Metal-poor BuLge stars with AAOmega survey (EMBLA) 
 (Howes et al. 2016) are ongoing.

Finally, by combining the accurate distance derived in this work, 
the Gaia proper motions, and the radial velocity from spectroscopy,
 we were able to compute a set of orbits for ESO456-SC38 in order to determine
 the most probable region where the cluster is moving in the Galaxy.
 The cluster has an eccentric orbit (e $>$ 0.7) that increases with the pattern speed of the Galactic bar,
with a radial distance between 0.1 and 2.5 kpc from the Galactic center
 and a maximum vertical excursion between 0.5 and 1.5 kpc.
 Its orbits are increasingly confined within the bulge and bar region, but even this does not support the bar structure, similarly to the old
 GCs HP~1 and NGC~6558.

\begin{acknowledgements}
SO and DN acknowledge partial support by the Universita` degli Studi
di Padova Progetto di Ateneo CPDA141214 and BIRD178590 and
by INAF under the program PRIN-INAF2014.
SOS acknowledges the FAPESP PhD fellowship no. 2018/22044-3.
 BB  and  EB acknowledge grants from  the
Brazilian  agencies  CAPES - Finance code 001, CNPq and  FAPESP. 
 APV acknowledges the FAPESP postdoctoral fellowship grant no. 2017/15893-1. 
\end{acknowledgements}


\end{document}